\documentclass[a4paper,12pt, reqno]{article}
\usepackage[latin1]{inputenc}
\usepackage{amsmath}
\usepackage{amsfonts}
\usepackage{amssymb}
\usepackage{graphicx}
\usepackage{subfigure}
\usepackage{caption}
\usepackage[usenames, dvipsnames]{color}

\def\ak{\theta_K}
\def\al{\theta_l}
\def\Sin{\text{sin}}
\def\Cos{\text{cos}}
\def\I{\text{I}}
\def\rat{\mathcal{R}}
\def\theprocess{$B\to K^* \ell^+\ell^-$}

\begin{document}
	\begin{center}
		
			\LARGE{Determining Form Factors and Wilson Coefficients using $B\rightarrow K^*\mu^+\mu^-$ data} \\ \vspace{.5cm}
		\large{ Bharti Kindra $^{a,b\star}$, Namit Mahajan$^{a\dagger}$} \\ \vspace{.5cm}
		
		\small{$\quad^a$ Physical Research Laboratory, Ahmedabad, India. \\
			$\quad^b$ Indian Institute of Technology, Gandhinagar, India. \\
			$\quad^\star$bharti@prl.res.in $\quad^\dagger$nmahajan@prl.res.in
		}
	\end{center}

	\textbf{Abstract:} A method to extract Wilson coefficients for semi-leptonic $b\rightarrow s$ transition and form factors for the hadronic process $B\rightarrow K^*\mu^+\mu^-$ is suggested. The method is based on data of angular distribution of the process $B\rightarrow K^*(\rightarrow K \pi)\mu^+\mu^-$ with the optimized observables plotted as functions of dilepton invariant mass squared. We have constructed plots of the observables using present knowledge of Wilson coefficients and form factors. Assuming these plots to be experimental data, we have discussed and used the strategy to determine the form factors and Wilson coefficients. The method is novel as this provides form-factor-independent determination of Wilson coefficients and also provides a reliable data driven method to extract information about form factors\\ 

\section{Introduction}
The rare decays induced by $b\to s \ell^+\ell^-$ transition are one of the potential probes of New Physics (NP). The process takes place through flavor changing neutral current transition which is loop suppressed as well as GIM suppressed exhibiting high sensitivity to New Physics contributions. The effective Hamiltonian for $b\to s\ell^+\ell^-$ transition is given by 

\begin{equation}
\mathcal{H}_{eff}=-\frac{4G_F}{\sqrt{2}} V_{tb}V_{ts}^* \mathcal{H}_{eff}^{(t)}+h.c.
\end{equation}
where, 

\begin{equation} \label{heff}
\mathcal{H}_{eff}^{(t)}=C_1 \mathcal{O}_1^c+ C_2\mathcal{O}_2^c +\sum_{i=3}^{6}C_i\mathcal{O}_i.
\end{equation}

The contribution due to up-quark is negligible and has been neglected, and the unitarity of the CKM matrix has been utilised to arrive at Eq.(\ref{heff}). Within the Standard Model, following operators have sizable contribution to the process \cite{symmetries},  

   \begin{eqnarray}\label{operators}
\mathcal{O}_7&=\frac{e}{g^2} m_b \left(\bar{s} \sigma_{\nu\mu}P_R b\right)F^{\mu\nu}, \nonumber \\
\mathcal{O}_9&=\frac{e^2}{g^2} \left(\bar{s}\gamma_{\mu}P_Lb\right)\left(\bar{\mu}\gamma^{\mu}\mu\right), \nonumber \\
\mathcal{O}_{10}&=\frac{e^2}{g^2} \left(\bar{s}\gamma_{\mu}P_Lb\right)\left(\bar{\mu}\gamma^{\mu}\gamma_5\mu\right)
\end{eqnarray}

In NP scenarios, operators other than the ones given in Eq.(\ref{heff}) can arise. Most prominent channel based on this transition is $B\to K^*\mu^+\mu^-$ which provides a multitude of observables through angular study of the decay which have been experimentally studied at LHCb, CMS, ATLAS, Belle and BABAR $ \cite{LHCb2017,LHCb2015,LHCb2014,LHCb2013,CMS2016,CMS2013,CDF2012,Belle2009,BaBar2006,BelleQ5}$. Several observables have shown deviations from predictions based on Standard Model, which includes $P^{\prime}_5$, Forward-Backward asymmetry ($A_{FB}$), branching ratio, Lepton-Flavour Universality Violating ratio $R_{K^{(*)}}$, and $Q_5=P^{\mu\prime}_5-P^{e\prime}_5$. Global analyses of these anomalies suggest a NP contribution to $b\to s \mu^+\mu^-$ through $C_9$ with/without additional contribution through $C_9^{\prime}$, $C_{10}^{(\prime)}$, and $\mathcal{O}_{T/T5}$ where primed Wilson coefficients corresponds to operators in Eq.(\ref{operators}) with helicity flipped \cite{matiasglobalfit,global1,global2,global3,Bardhan}.\newline 

  Other than Wilson coefficients, these observables depend on various hadronic corrections. As discussed in several works \cite{charmloopNP,camalich,ccloopciuchini,ccloophurth}, charm loop corrections contribute to \theprocess process through $C_9$ which can be a potential source of the deviations observed. On the other hand, the deviations in $R_K$ and $R_{K^*}$ are a telltale signs of NP as the hadronic contribution in muon channel cancels (almost) the hadronic contribution of electron channel \cite{Hillerrkrkstar}. But these deviations are not statistically significant yet ($\approx 2.5 \sigma$). Thus, to understand the nature of these deviations, it is essential to isolate the NP effects at short-distance from that of hadronic uncertainties in the long-distance Standard Model contributions. Several attempts in this direction have been made   \cite{MahmoudiWC,DasSinha1,DasSinha2,Rusa1,Rusa2}, where either experimental techniques or theoretical expressions have been discussed as an absolute test of NP. In \cite{Khodjamirianff}, efforts have been made to account for the hadronic contribution systematically for $B\to K\ell^+\ell^-$. \newline
  
 In this work, a model independent technique has been suggested to obtain Wilson coefficients and form factors from the angular analysis of $B\rightarrow K^*\mu^+\mu^-$ data. This method requires the data to be plotted as a function of dilepton invariant mass squared ($q^2$). When plotted, many observables intersect each other or cross the $q^2$ axis. The value of $q^2$ where an observable vanishes is called a zero crossing. The zero crossing of $A_{FB}$ is long known to be reasonably independent of the hadronic uncertainties \cite{afbzerobeneke,afbzeroburdman}. It has been pointed out in literature that the zero crossings of $P_4',P_5',$ $A_{FB}$ and $O_T$ provide correlations between Wilson coefficients which are independent of the model under study as well as hadronic parameters \cite{Alifbzero,girish}. On the experimental side, the value of zero-crossing of $A_{FB}$ has also been reported by LHCb to be $4.9\pm 0.9$ Gev$^2$( $\approx 20\%$ uncertainty) \cite{LHCbzero}. In this work, we discuss the strategy to extract values of the Wilson coefficients using these zero crossings. Other crossings, where two observables intersect each other are also utilized to extract Wilson coefficients as well as form factors.  \newline
      
   The paper is structured as follows. In Section 2, all the observables have been discussed and are plotted as a function of $q^2$. We then take these plots as psuedo-experimental data and discuss the strategy to extract the values of form factors and Wilson coefficients using this pseudo-experimental data in Section 3. We finally conclude and discuss the prospects in Section 4.

      \section{Observables as function of $q^2$}
      
   Experimental input of $B\rightarrow K^*\mu^+\mu^-$ is based on the four body differential decay distribution of $ K^*(\rightarrow K \pi )\mu^+\mu^-$. It is expressed in terms of four kinematic variables ($\{q^2,\theta_K,\theta_l,\phi\}$) as \cite{completeanatomy}, 
   
      \begin{equation}\label{differential}
      \begin{split}
      \frac{d^4\Gamma}{dq^2 ~d\Cos\theta_K~d\Cos\theta_l~d\phi}=& \frac{9}{32 \pi}[\I_{1s} \Sin^2\ak +\I_{1c} \Cos^2\ak+ (\I_{2s}\Sin^2\ak+\I_{2c}\Cos^2\ak) \Cos 2\al  \\ & + \I_3 \Sin^2\ak~\Sin^2\al~\Cos 2 \phi+\I_4\Sin 2\ak~\Sin 2\al~cos\phi \\ & + \I_5 \Sin 2\ak~sin\al~\Cos\phi+(\I_{6s} \Sin^2~\ak +\I_{6c}\Cos\ak)\Cos\al \\ & \I_7 \Sin 2\ak~\Sin\al~\Sin\phi +\I_8\Sin 2\ak ~\Sin 2\al ~\Sin\phi \\ & \I_9 \Sin^2\ak~\Sin^2\al~\Sin 2\phi  ] 
      \end{split}
      \end{equation}  
      The angular coefficients $\I_i(q^2)$ can be expressed in terms of transversity amplitudes ($A_i$), which in the effective Hamiltonin approach, depend on Wilson coefficients which encode short-distance effects as well as potential new physics and form factors which contain the hadronic contributions and are source of large theoretical uncertainties. Explicit expressions for $\I_i(q^2)$ and description of the kinematic variables are given in the Appendix A. \newline
      
      There are seven non-perturbative form factors that describe the amplitude of $B\rightarrow K^*\mu^+\mu^-$. These are evaluated using Light Cone Sum Rules (LCSR) \cite{QCDsumrules,BallZwicky} which are based on certain assumptions that introduce systematic uncertainties. In heavy quark limit ($m_b\rightarrow \infty$) and large recoil limit ($q^2\rightarrow 0$), these seven form factors reduce to two independent form factors ($\xi_{\perp},\xi_{\parallel}$) called soft form factors \cite{Charles}. However, there are subleading terms of the order of  $\Lambda_{QCD}/m_b$ and $\alpha_s$ (strong coupling constant). For systematic study, following parameterization is often used to write form factors \cite{stateoftheart}.
      \begin{equation}\label{decompoosition}
      F(q^2)=F^{\infty}(\xi_{\perp}(q^2),\xi_{\parallel}(q^2))+\Delta F^{\alpha_s}(q^2)+\Delta F^{\Lambda}(q^2)
      \end{equation}
      where, $\Delta F^{\alpha_s}$ are corrections ($\mathcal{O}(\alpha_s)$) due to the hard gluon exchanges while $\Delta F^{\Lambda}$ are corrections ($\mathcal{O}(\Lambda_{QCD}/m_b)$) due to the soft gluon exchanges. These are called factorisable corrections as they contribute to the form factors. There are other corrections that are not related to form factors, like charm loop corrections, called non-factorisable corrections. These corrections are not well computed but a naive estimate of all these corrections amounts to $25-30 \%$ of errors in form factors.\newline
      
\begin{figure}[h]
      	\begin{center}
      		\includegraphics[scale=0.6]{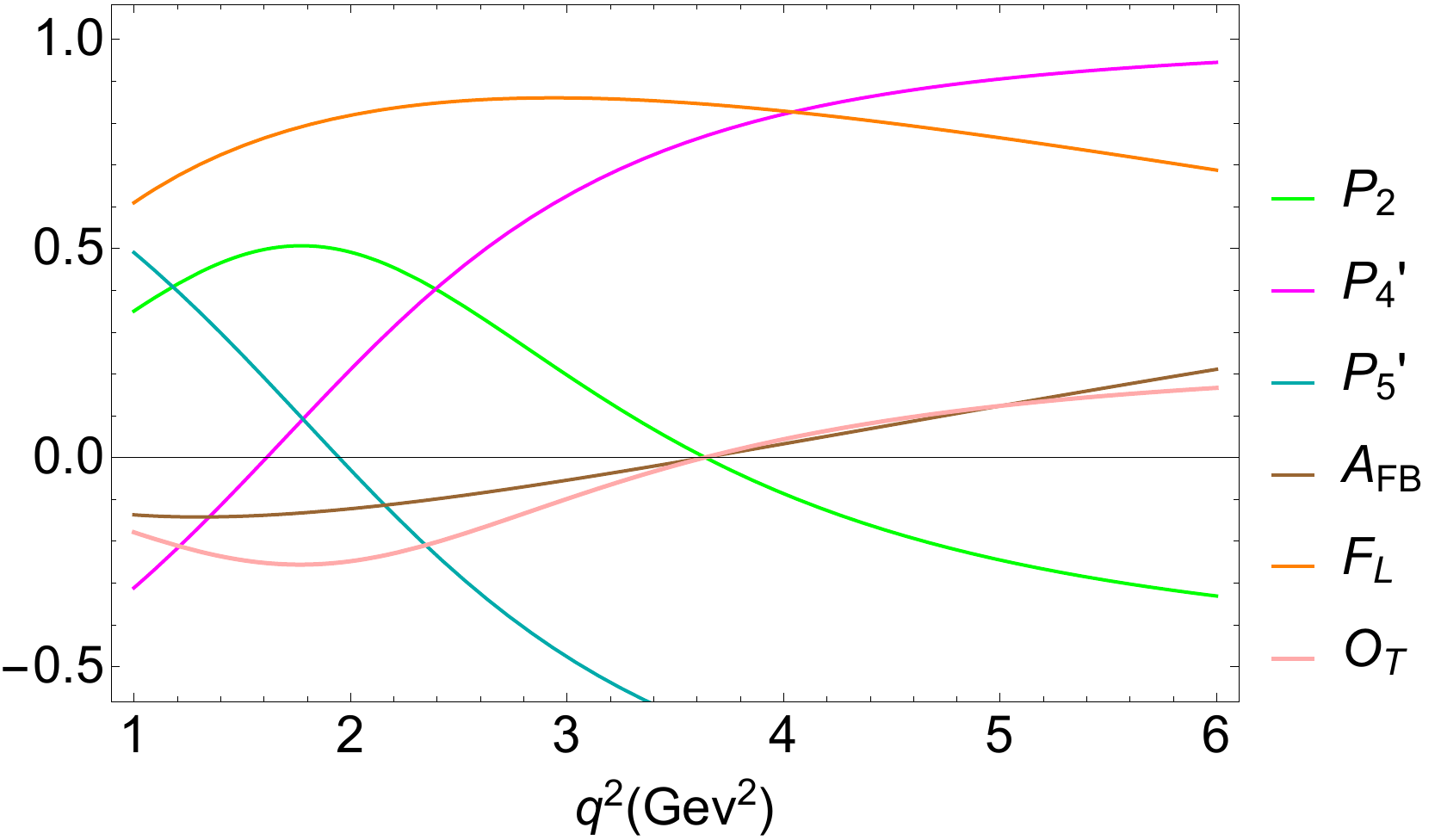}
      	\caption{Relevant observables as a function of $q^2$. \label{crossings}}
      	\end{center}
      \end{figure}

      The decomposition of form factors as shown in Eq.(\ref{decompoosition}), has been a motivation for construction of optimized observables \cite{optimizing}. In such observables, dependence of soft form factors cancels atleast in the leading order, reducing the hadronic uncertainties in the predictions. The definitions of observables used are, 

      \begin{align}
      P_1&=\frac{1}{2}\frac{\I_3+\bar{\I}_3}{J_{2s}+\bar{\I}_{2s}} & P_2&=\frac{1}{8}\frac{\I_{6s}+\bar{\I}_{6s}}{\I_{2s}+\bar{J}_{2s}}  \\
      P_3&=-\frac{1}{4}\frac{\I_9+\bar{\I}_9}{\I_{2s}+\bar{\I}_{2s}} & P_4'&=\frac{1}{\mathcal{N}}(\I_4+\bar{\I}_4)  \\
      P_5'&=\frac{1}{2 \mathcal{N}}(\I_5+\bar{\I}_5) & 
      P_6'&=-\frac{1}{2\mathcal{N}}(\I_7+\bar{\I}_7)\\ 
      A_{FB}&=-\frac{3}{4} \frac{\I_6+\bar{\I}_{6}}{d\Gamma/dq^2+d\bar{\Gamma}/dq^2} &    F_L&=-\frac{\I_{2c}+\bar{\I}_{2c}}{d\Gamma/dq^2+d\bar{\Gamma}/dq^2}\\
      \frac{dBR}{dq^2}&=\tau_B \frac{d\Gamma/dq^2+d\bar{\Gamma}/dq^2}{2} & O_T &= \frac{\left|A_{\perp}^L\right|^2+\left|A_{\parallel}^L\right|^2-(L\leftrightarrow R)}{8(\I_{2s}+\bar{\I}_{2s})}
      \end{align}
      where, $$\mathcal{N}=\sqrt{-(\I_{2s}+\bar{\I}_{2s})(\I_{2c}+\bar{\I}_{2c})},$$ $$d\Gamma/dq^2=\frac{1}{4}(\I_{1c}+6 \I_{1s}-\I_{2c}-2\I_{2s}),$$
       $A_{FB}$ is the forward-backward asymmetry, $F_L$ is the longitudinal polarization fraction, $dBR/dq^2$ is the differential branching ratio, and $\tau_B$ is the life time of B meson. $\bar{\I}_i(q^2)$ are CP-conjugate functions of angular coefficients defined in Eq.(\ref{differential}) obtained by replacing  
      \begin{align}
      \I_{1,2,3,4,7}&\rightarrow  \bar{\I}_{1,2,3,4,7} & \I_{5,6,8,9}\rightarrow -\bar{\I}_{5,6,8,9}
      \end{align}
      These observables are shown as functions of $q^2$ in Fig. \ref{crossings}. Since the aim of this work is to outline a strategy and not precision, we are including only the central values of form factors and other input parameters to generate the plots in the Fig. \ref{crossings}.  While discussing the strategy in the next section, we include a deviation of $5\%$ and $10\%$ in the value of crossings to account for the experimental error. Also, our plots may differ a bit from other theoretical predictions \cite{alpha2,alizeroprediction,girish} as the crossings are sensitive to the input parameters. Form factors, values of masses and couplings, and Wilson coefficients used here are given in the Appendix \ref{input}. We are not concerned about the rigor here and use Fig. \ref{crossings} to portray experimentally obtained plots and discuss the strategy to extract form factors and Wilson coefficients using this pseudo-experimental information.  \newline
      
      It can be noted that there are four classes of crossings in Fig. \ref{crossings}: \newline \newline
      Class 1: $q^2$ value, where an observable which is independent of form factors vanishes (or crosses the $q^2$ axis).\\ 
      	 Class 2: $q^2$ value, where an observable which is dependent of form factors vanishes.\\ 
      	Class 3: $q^2$ value, where two observables having same dependence on form factors crossing each other.\\      	
      	Class 4: $q^2$ value, where two observables having different dependence on form factors crossing each other.\\ 
      	     
        We refer to the first two class of crossings as \textit{zero crossings} and the latter two as \textit{observable crossings}. In the next section, we show how to use each of these crossings to extract information about the input parameters.
      
      \section{Strategy} 
      
      The zero crossing of $A_{FB}$, which lies in theoretically clean low $q^2$ region ($q^2<6$Gev$^2$), has been measured by LHCb which an error of approximately 20$\%$. Theoretically, within Standard Model, this zero crossing in the leading order,  determines the ratio of $C_7$ and $C_9$. Other observables including $P_2$, $O_T$, $P_4^{\prime}$, and $P_5^{\prime}$ also have their zeros in low $q^2$ range and provide relations between different Wilson coefficients. The advantage of using zero crossings is two fold. First, they provide simple relations between Wilson coefficients which are form factors and model independent.\footnote{We restrict our operator basis to that discussed in Eq.(\ref{heff}). Generalization to a larger set is straightforward and the relations may even change significantly. Also, we assume Wilson coefficients to be real for the present study.} Second, as discussed for $A_{FB}$ in \cite{afb}, form-factor-dependence vanishes at these zero crossings in the leading order resulting in less theoretical error. This means that even small deviations, if found experimentally will constitute a sign of NP. However, these crossings do not provide complete information about Wilson coefficients and form factors. For example, zero crossings determine the values of $C_9$ and $\left|C_{10}\right|$ but not the sign of $C_{10}$. Therefore, despite being not as clean as zero crossings, observable crossings will play an essential role in this work. As shown later in this section, Class 3 observables provide relations sensitive to sign($C_{10}$). $\left|C_7\right|$ is well determined from $ B \rightarrow K^* \gamma$. Using that as an input, $C_9$ and $C_{10}$ are calculated using equalities in Table~\ref{type1} and Table \ref{type3}. \\
      
      Once the Wilson coefficients are obtained, they are substituted in relations obtained from Class 4 observables, which depend on the ratio of soft form factors. Experimental determination of branching ratio then allows to get the values of both soft form factors separately.
      
      \subsection{Extraction of Wilson Coefficients}
      Class 1 and Class 3 crossings, by definition, are independent of form factors and provide relations between different Wilson coefficients. Also, form factor dependence cancels out at the zero crossing of Class 2 observables. An example of this Class is $A_{FB}$. Therefore, these are also used to extract Wilson coefficients. No numerical values are determined at this stage and we show only the relations that can be used to do so. \newline 
      
      \begin{table}[h]
      	\begin{center}
      		\begin{tabular}{|c|c|c|}
      			\hline	Observable& $s_0$(Gev$^2$) & Relation between Wilson coefficients\\  \hline
      			$P_2$& 3.52& E1=$(2 C_7 \hat{m_b}+C_9 \hat{s})=0$  \\
      			$P_4'$& 1.55&E2=$((C_9^2+C_{10}^2)\hat{s}+4C_7^2\hat{m}_b^2+2C_7C_9\hat{m}_b(1+\hat{s}))=0$\\
      			$P_5'$ & 1.8& E3=$(C_9\hat{s}+C_7\hat{m}_b (1+\hat{s}))=0$\\
      			$O_T$ & 3.52 &  E4=E1=($2 C_7 \hat{m_b}+C_9 \hat{s})=0$ \\  \hline
      		\end{tabular}
      		\caption{Class 1 crossings and corresponding relations between Wilson coefficients.} \label{type1}
      	\end{center}
      \end{table}
      
      In Table \ref{type1}, Class 1 crossings have been listed along with the zero ($q^2$) values at which the corresponding relation between Wilson coefficient follows.\footnote{Again, these relations are valid as long as we are restricted to operator basis in Eq.(\ref{heff})} The expressions on the left hand side of the equation have been labeled as E1; E2; E3. This enables us 
      to correlate expressions derived for class 3 observables in terms of those for Class 1. As an illustration, consider the observable $P_2$. It vanishes when $E1= 2 C_7\hat{m}_b+C_9\hat{s}$, vanishes at $q^2_0=3.52$Gev$^2$.\footnote{ Throughout the paper, the hatted quantities denote the corresponding quantities normalized to $m_B$ appropriately to render the hatted quantities dimensionless and $s\equiv q^2$.} This yields $\hat{s}_0(P_2)=-2 (C_7/C_9)\hat{m}_b$ which is the celebrated zero crossing of $A_{FB}$ as well. Such a relation and the corresponding measurement of zero crossing points yield further information about Wilson coefficients. In the further analysis, we can thus rewrite the ratio $(C_7/C_9)$ in terms of $\hat{s}_0(P_2)=\hat{s}_0(A_{FB})$ which is an exprimentally measured quantity. We follow this approach for other observables as well.\newline
      
      
      
      In Table \ref{type3}, observable crossings and corresponding values of $q^2$ have been shown where, $\beta=(1-4(m_{\mu}^2/q^2))^{1/2}$ is the phase space factor. When two observables cross each other at a specific $q^2$, denoted as $s_c$, the two observables at the point are equated and this yields relations between Wilson coefficients in terms of experimentally measured or measurable quantities. The corresponding relation between Wilson coefficients at these crossings have been expressed in terms of zero values of observables given in Table \ref{type1} and $\hat{s}_0$ of $A_{FB}$ i.e, $\hat{s}_0=-2 \frac{C_7}{C_9}\hat{m}_b$. Matching these expressions with the actual experimentally obtained value of crossing can be used to extract Wilson coefficients $C_9$ and $C_{10}$ in a model independent way.
      \begin{table}[h]
      	\begin{center}
      		\begin{tabular}{|c|c|c|}
      			\hline
      			Observables & $s_c$(Gev$^2$)& Functional Form \\ \hline 
      			&&\\
      			$P_5', P_2$& 1.13& $\frac{E3}{\left[C_9^2(1+\hat{s}_0^2-2\hat{s}_0)+C_{10}^2\right]^{1/2}}=\frac{E1}{\left[C_9^2(1+\hat{s}_0^2/\hat{s}^2-2\hat{s}_0/\hat{s})+C_{10}^2\right]^{1/2}}$\\ &&\\
      			$P_5',P_4'$& 1.71 & $\frac{2 C_{10}E3}{\hat{s}\left(C_9^2(1+\hat{s}_0^2/\hat{s}-\hat{s}_0/\hat{s}+C_{10}^2\right)}=\beta$\\ &&\\
      			$P_4',P_2$& 2.31& $\frac{\left[C_9^2(1+\hat{s}_0^2-2\hat{s}_0)+C_{10}^2\right]^{1/2}}{\left[C_9^2(1+\hat{s}_0^2/\hat{s}^2-2\hat{s}_0/\hat{s})+C_{10}^2\right]^{1/2}} \frac{(C_{10} E1) }{E2}=\beta$ \\ \hline
      		\end{tabular}
      		\caption{Class 3 crossings and corresponding relations between Wilson coefficients} \label{type3}	
      	\end{center}
      \end{table}

     Note that Table \ref{type1} and \ref{type3} can determine only the central values of Wilson coefficients. The error bands can be systematically added using the uncertainties in measurement of zero crossings. Taking the values of Wilson coefficients determined using this strategy as input values, we next discuss the procedure of extracting the form factors.

      \subsection{Extraction of form factors}
      
      Class 4 are the crossings that depend on form factors i.e, when the two observables are mathematically equated to each other at the point where they intersect, there is a form factor dependence that remains in the functional form. These crossings are shown in Table \ref{type4}.   
      \begin{table}[h]
      	\begin{center}
      		\begin{tabular}{|c|c|} 
      			\hline 
      			Observables & $s_*$(Gev$^2$)\\ \hline 
      			$A_{FB}, P_4^{\prime}$ & 1.35 \\ 
      			$A_{FB}, P_5^{\prime}$ & 2.16 \\
      			$F_L , P_4^{\prime} $ & 4.04 \\ 
      				$A_{FB},O_T $ & 5.00 \\ \hline 
      		\end{tabular}
      		\caption{Class 4 crossings and corresponding $q^2$ values.} \label{type4}
      		
      	\end{center} 
      \end{table}
  Given that the Wilson coefficients have been fixed using the procedure discussed in the previous section, all of these crossings can be expressed as functions of the ratio $\rat= \xi_{\perp}/\xi_{\parallel}$ and the point ($q^2$) where the two observables cross. Thus, determining the crossing point experimentally will fix the ratio of form factors at $s_*$ i.e, a given value of $s_*$ allows to extract ratio at that particular $s_*$ ($\rat(s_*)=\xi_{\perp}(s_*)/\xi_{\parallel}(s_*)$). Utilizing various such $s_*$ values, we can have a handle on $\rat(q^2)$ i.e, we start to get the functional information of the ratio $\rat$. \\
        \begin{figure}[h]
   	\begin{center}
   		\begin{tabular}{c c}
   			\includegraphics[scale=0.5]{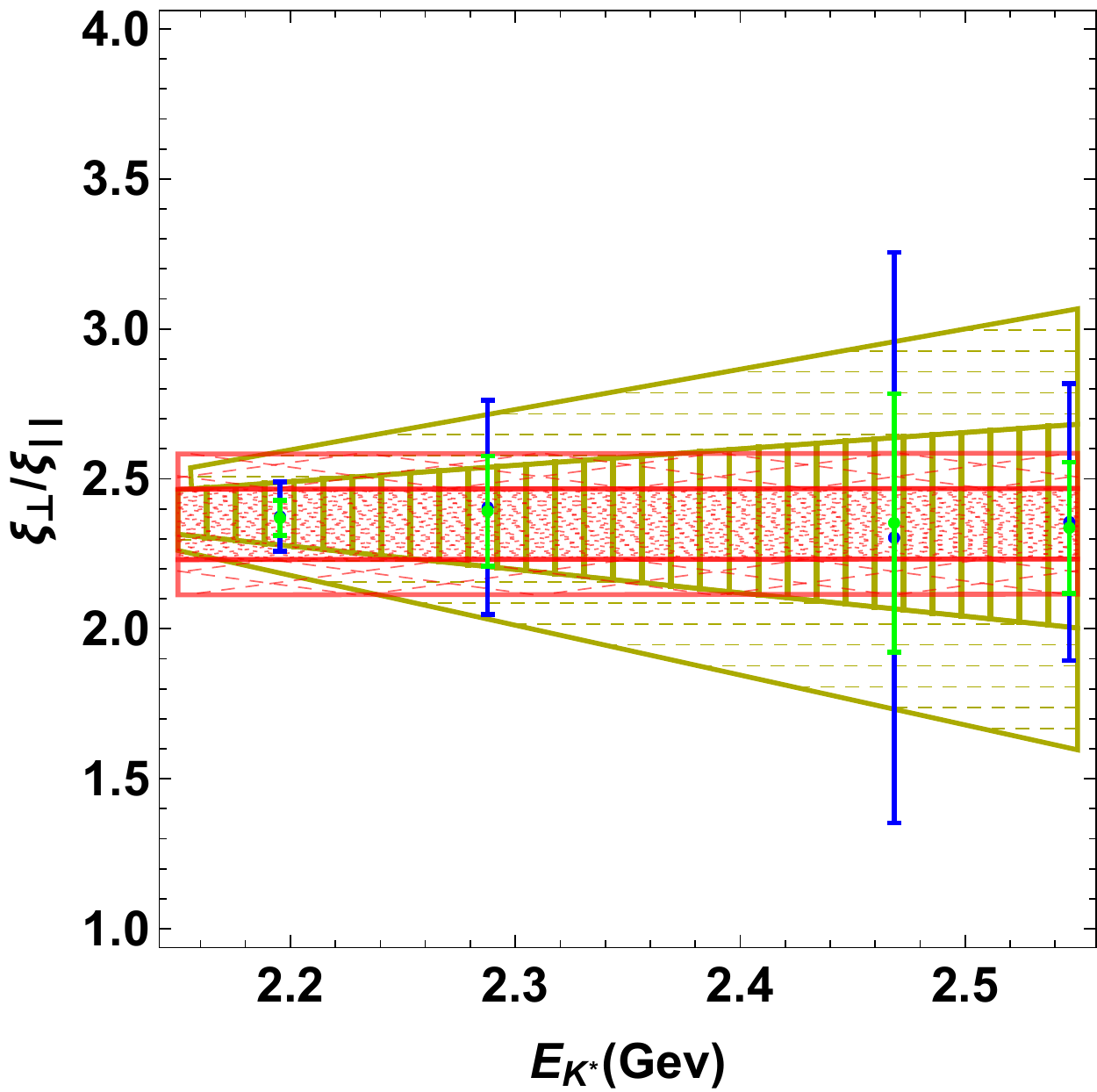}& \includegraphics[scale=0.5]{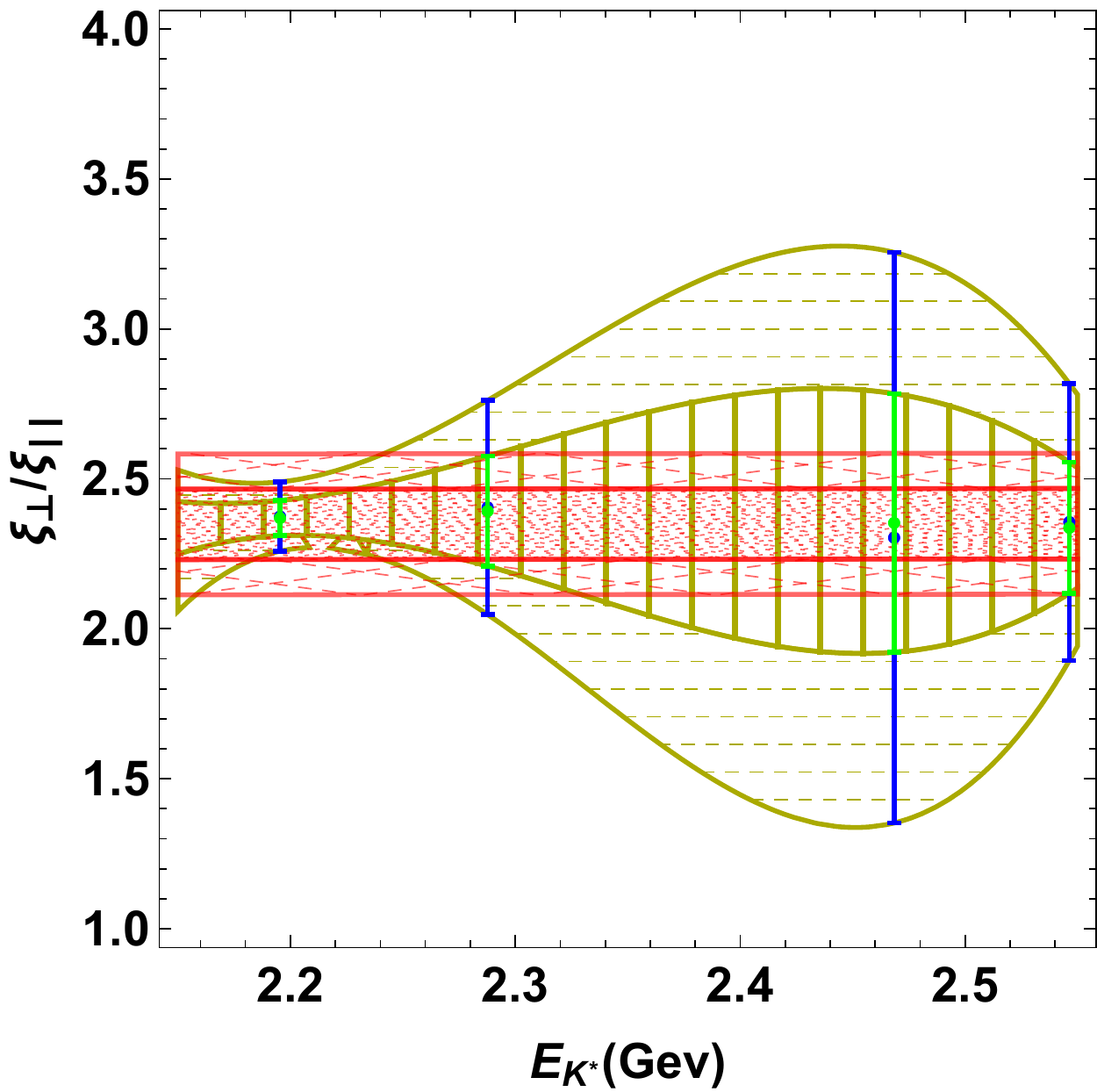}\\
   			\textbf{(a)}
   			\vspace{2mm}&
   			\textbf{(b)}
   			\vspace{2mm}\\
   		\end{tabular}
   		\caption{(a) Linear fit (b) Cubic fit. Solid and Dashed Yellow respectively show the extracted values with $5\%$ and $10\%$ errors assumed in measurement of crossing points. Dotted and Dashed Red respectively show $5\%$ and $10\%$ errors in the form factors used to generate Fig. \ref{crossings}. }
   		\label{ratios}
   	\end{center}
   \end{figure}
   
   As discussed before, Fig. \ref{crossings} shows various observables as a function of $q^2(=s)$(Gev$^2$), drawn for Standard Model and employing only the central values of Wilson coefficients and form factors. To account for the statistical error in the determination of crossing points, we assume an uncertainty of $10\%$ in the value of crossings. To illustrate, we consider a crossing point, say $s_1$ and assume that it is known with an error of $10\%$ i.e., the point lies somewhere in the interval $\left[0.9 s_*,1.1s_* \right]$. In this interval, we generate 1000 random points. At one of these points, say $s_1$, we equate the two observables under consideration and get the value of ratio, $r_1$. Mean and variance of ratios calculated at these 1000 points yield $\rat\pm \delta\rat$ corresponding to the considered point $s_*$. \newline

   We repeat the procedure for $5\%$ uncertainty in the determination of crossing point. The extracted values of $\rat\pm\delta\rat$ obtained using this strategy are shown in Fig. \ref{ratios}. These are only representative values which we use to show how the method works and how $5\%$ determination can help to extract form factors with small errors. Using the Class 4 crossing points (there are four of these points spanning $q^2$ between 1.3-5 Gev$^2$), the ratio $\rat$ can be extracted at these four points. We then employ two representative fits, linear and cubic, to get a reasonable functional form of $\rat(q^2)$. Since in the functional form of all the observables ($O_i(q^2)$), soft form factors enter as a function of energy carried by the light hadron ($E_{K^*}=(m_B^2+m_{K^*}^2-q^2)/(2 m_B)$), we have plotted the form factors as a function of $E_{K^*}$. \newline
   
     \begin{figure}[h]
    	\centering
    	\begin{tabular}{cc}
    		\includegraphics[scale=0.5]{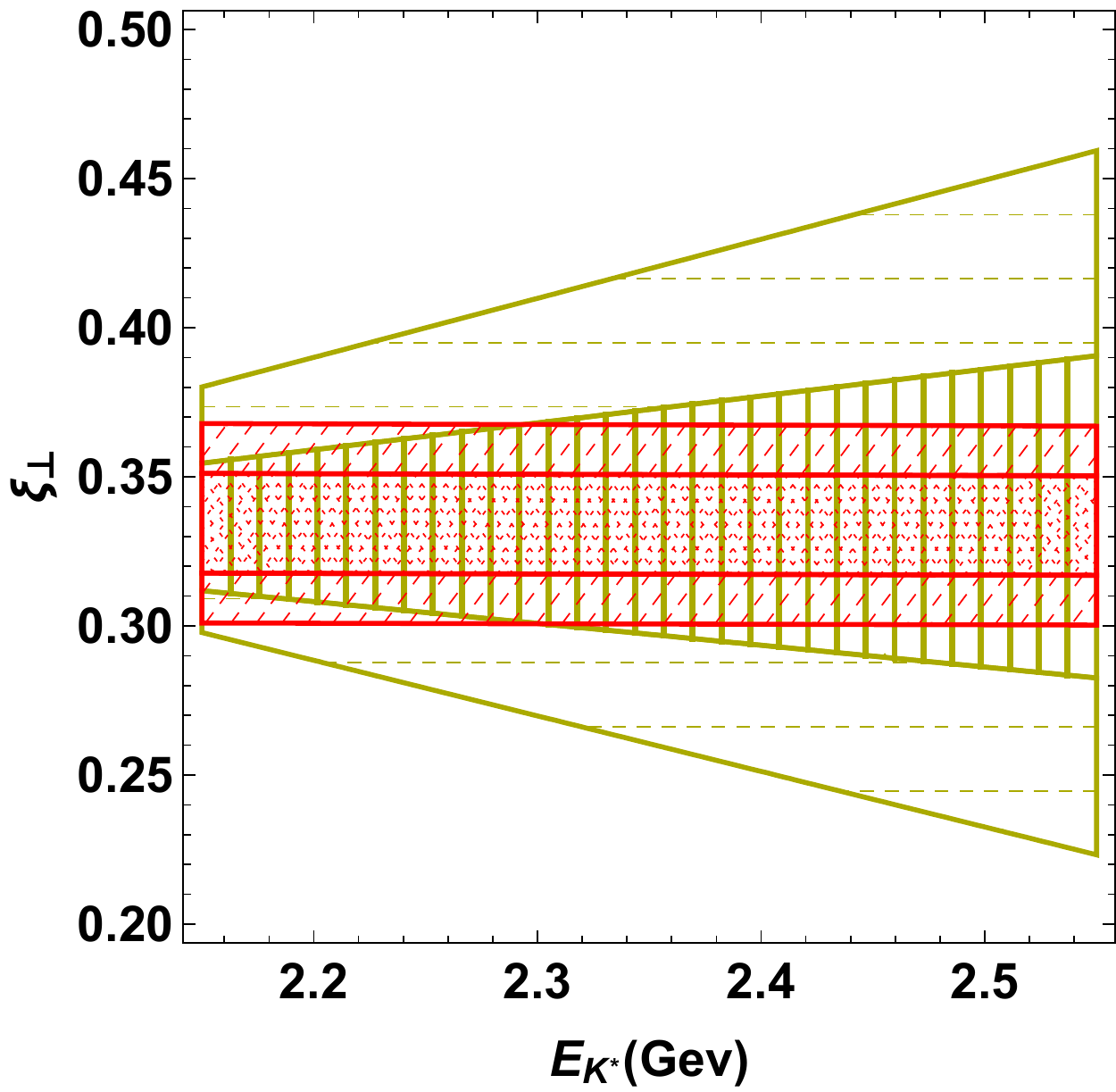}& \includegraphics[scale=0.5]{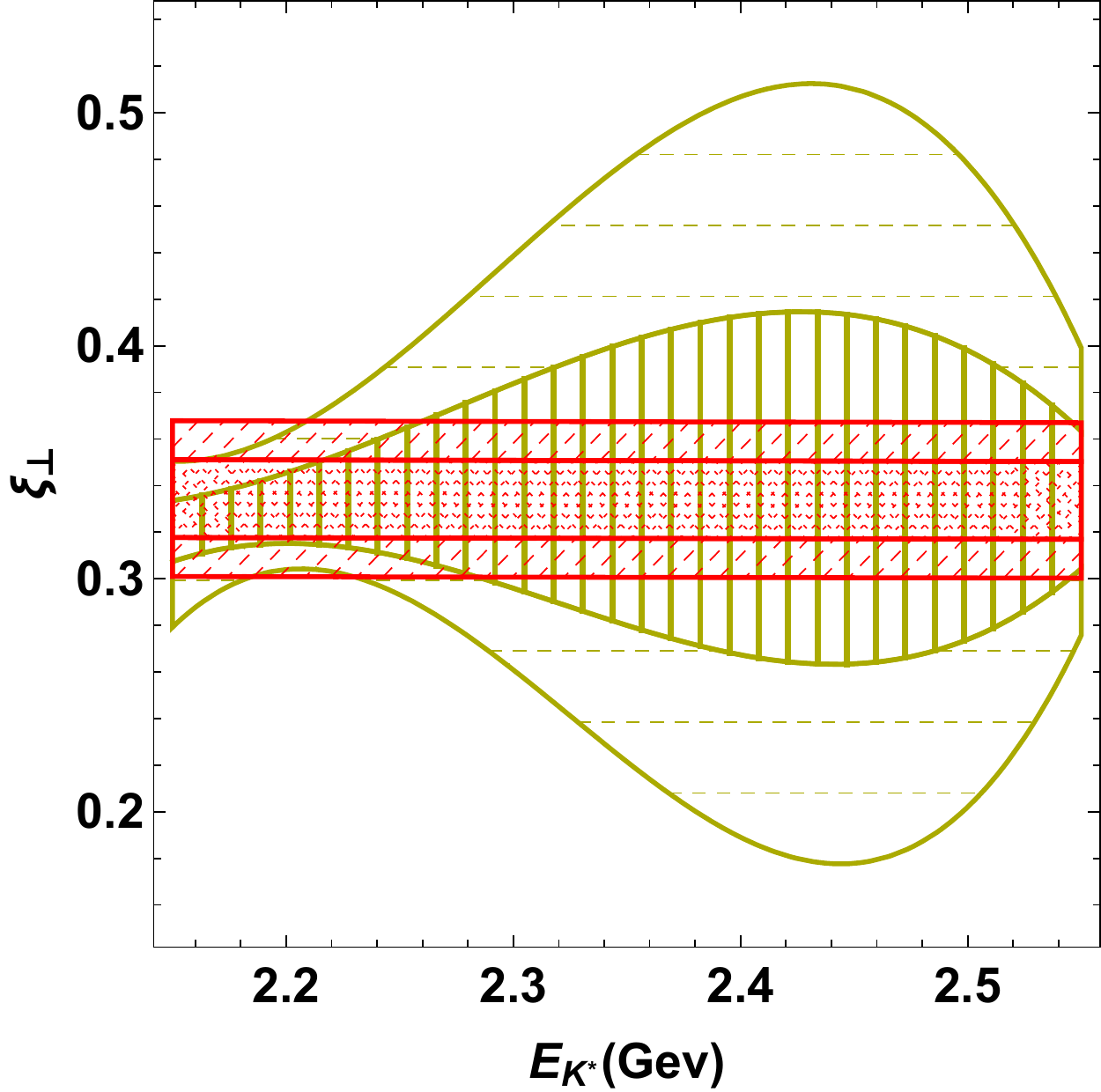}\\
    		\textbf{(a)}
    		\vspace{2mm} &      	
    		\textbf{(b)}
    		\vspace{2mm}\\
    	\end{tabular} 
    	\caption{  \label{perp} (a) Linear fit (b) Cubic fit. Solid and Dashed Yellow respectively show the extracted values with $5\%$ and $10\%$ errors assumed in measurement fo crossing points. Dotted and Dashed Red respectively show $5\%$ and $10\%$ errors in the form factors used to generate Fig. \ref{crossings}.  }
    \end{figure}
    
    \begin{figure}[h]
    	\centering
    	\begin{tabular}{cc} 
    		\includegraphics[scale=0.5]{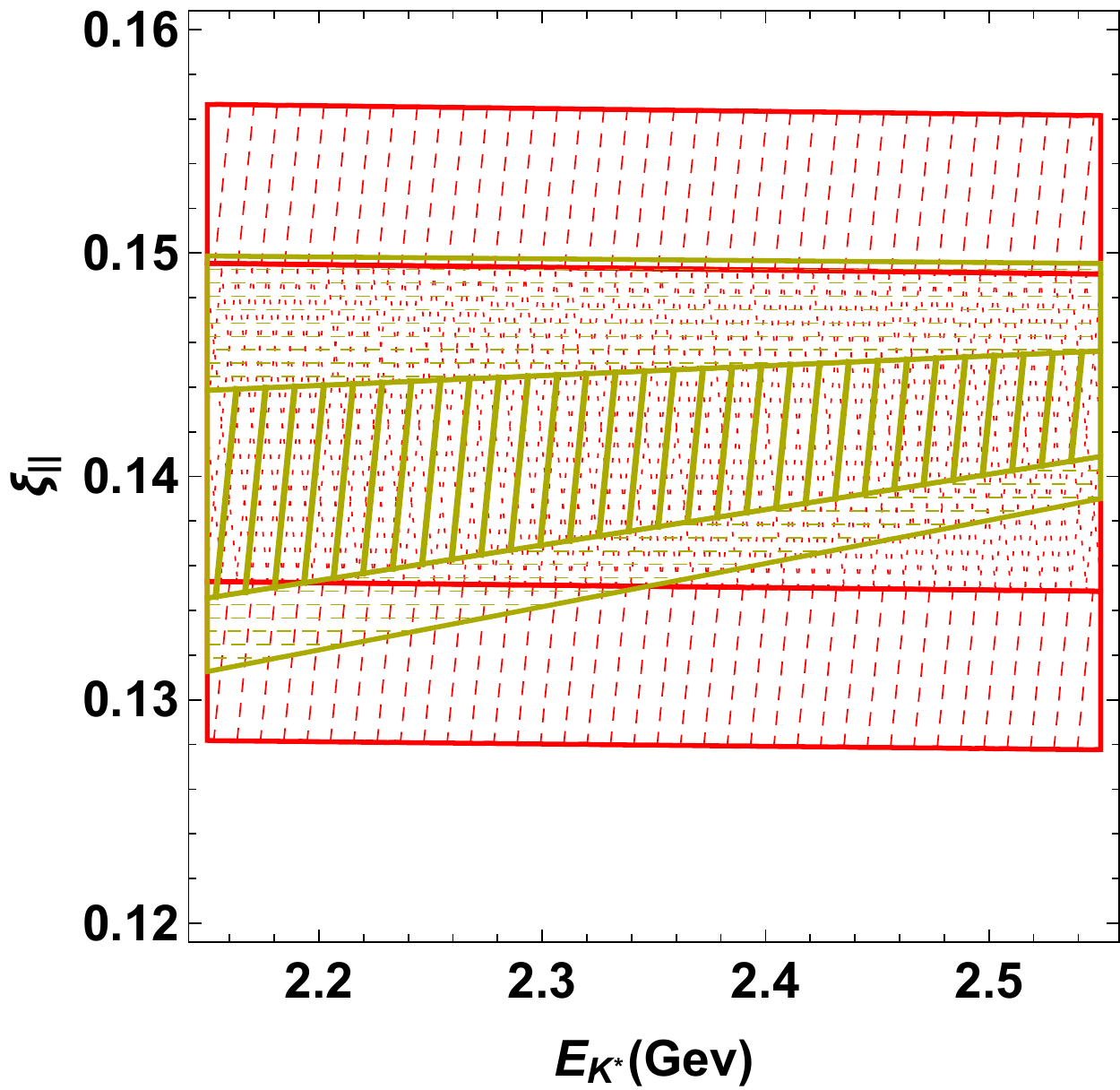}& \includegraphics[scale=0.5]{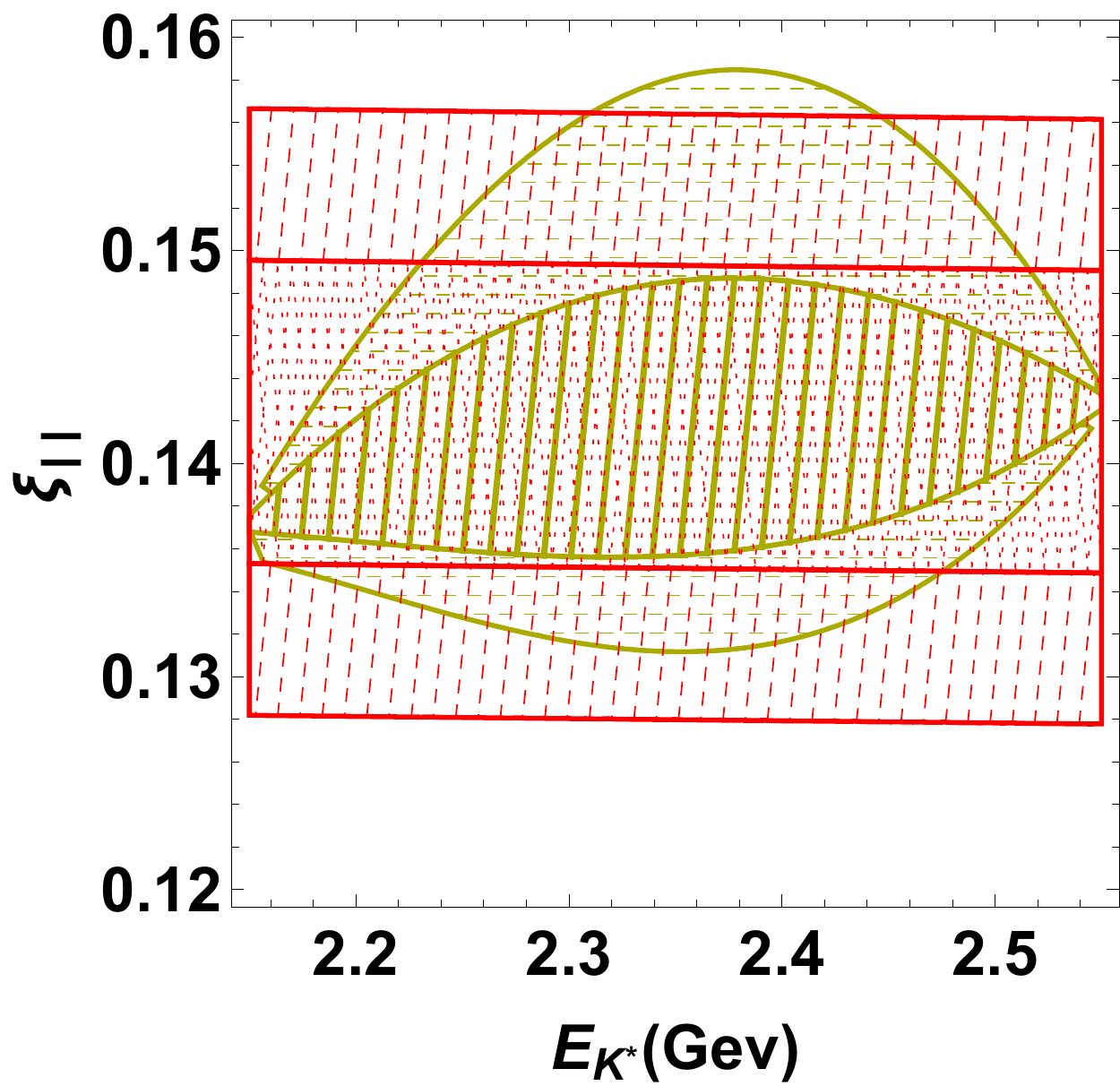}\\
    		\textbf{(a)}
    		\vspace{2mm}&      	
    		\textbf{(b)}
    		\vspace{2mm}\\
    	\end{tabular}
    	\caption{\label{para} (a) Linear fit (b) Cubic fit. Solid and Dashed Yellow respectively show the extracted values with $5\%$ and $10\%$ errors assumed in measurement fo crossing points. Dotted and Dashed Red respectively show $5\%$ and $10\%$ errors in the form factors used to generate Fig. \ref{crossings}.}
    \end{figure}
      To find the values of each of the soft form factors, any observable which does not belong to the list of optimized observable can be used. To demonstrate the procedure we use branching ratio which depends on both $\xi_{\parallel}$ and $\xi_{\perp}$ but can easily be recast to depend on say $\xi_{\perp}$ and, $\rat$ which as just discussed, is determined (atleast in principle at this stage) employing Class 4 crossings. Thus, a functional fit to $dBR/dq^2$ with $\rat(q^2)$, determined above, taken as the input effectively determines $\xi_{\perp}(q^2)$. Determination of $\xi_{\parallel}(q^2)$ from $\rat(q^2)$ and $\xi_{\perp}(q^2)$ is straight forward. The extracted plots are shown in Fig. \ref{perp} and \ref{para} with an assumed $10\%$ error (Yellow-dashed) and a more optimistic scenario with $5\%$ error (Yellow-Solid) in various crossings. Also shown for comparison are the bands (Red) for form factors and their ratio which was used to get the plots in Fig. \ref{crossings} (again we have shown $5\%$ and $10\%$ errors here).\newline 
      
      The idea is to pretend that these theoretical bands are the true Form Factor bands, and to check if a data driven strategy is able to yield bands that come close to these \textit{"true bands"}. We see that a $5\%$ determination of the crossing points results in a Form Factor determination that is very close to the true bands.

      \section{Conclusion}
      Semi-leptonic $B \to K^*\ell^+\ell^-$ decays provide a unique window to possible physics beyond the Standard Model. These relatively cleaner modes still suffer from significant hadronic uncertainties, stemming from form factors. In recent times, several anomalies have been reported which possibly are a sign of new physics. However, the hadronic uncertainties do not allow one to unambiguously reach this conclusion. It is highly necessary to disentangle the effects that may be arising due to our lack of knowledge of hadronic contributions. A data driven strategy is therefore a good option, provided these hadronic effects are kept in control. In this paper, we have suggested such a method. \newline
      
      The central point of the method is to make optimum utilization of various crossing points, zero crossings or crossing points of two observables. These points become useful due to the fact that not only there are correlations among different observables and crossing points, these crossing points are often (almost) independent of the form factors, and thus of the hadronic uncertainties. Thus, enabling an extraction of the Wilson coefficients which encode the useful new physics information. There are other crossing points that can be utilized, in association with the information already extracted, to determine the functional form of the form factors. As we have shown, the method is enormously powerful and the only limitation seems to be the accuracy with which these crossing points can be determined experimentally. Assuming a good determination ($5\%$ accuracy), we have shown that the form factors can actually be extracted to a good precision. Such a data driven strategy will help in eliminating any bias towards theoretical inputs. For example, while performing global fits, one has to assume certain form factor behavior, which may bias the fits. We strongly believe that dedicated efforts in measuring the crossing points will really help in establishing the presence of new physics, if present. Moreover, the method can be used to make these decays a QCD laboratory to test our theoretical understanding of form factors. For example, a similar strategy could be used for the mode $B_s\to \phi \mu^+\mu^-$. A direct comparison of the form factors extracted for $B_s\to \phi \mu^+\mu^-$ and $B_d \to K^*\mu^+\mu^-$ can then be used to study $SU(3)$ effects. \newline
      
      We hope that in the near future, the crossing points will be measured with better accuracy and the method suggested here (or a more refined version) can be utilized to determine both the form factors as well as Wilson coefficients in a model independent way. Also, we note that a linear fit to the "\textit{experimental data}" gives a better determination. 
      \appendix
      \section{Appendix}
      \subsection{Kinematic variables}\label{kinematicvariables}
\begin{figure}[h]
      	\centering
      	\includegraphics[scale=0.6]{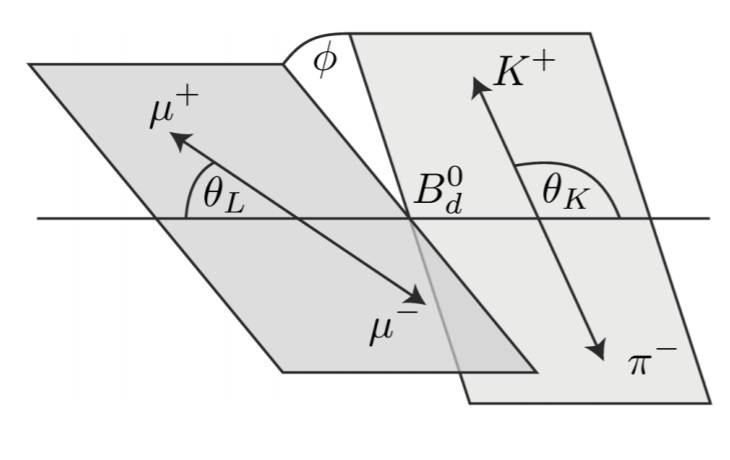}
      	
      	\caption{The kinematics of 4-body decay\label{kinematics}}
      \end{figure}

      The differential decay distribution of the four body decay $B\rightarrow K^*(\rightarrow ~K\pi)\ell^+\ell^-$ is defined is terms of four parameters (Fig. \ref{kinematics}): 
      \begin{itemize}
      	\item $\theta_K \in [0,\pi]$, the angle that the resultant Kaon makes with the B meson in the rest frame of $K^*$; 
      	\item $\theta_l\in [0,\pi]$, the angle between flight of $\ell^-$ and the B meson in the dilepton rest frame ;
      	\item $\phi \in [-\pi,\pi]$, the azimuthal angle between the two planes defined by the lepton pair and the $K \pi $ system;
      	\item $q^2$, the invariant squared mass of the lepton pair. 
      \end{itemize}
      
      \subsection{Angular Coefficients}\label{angular}
      Th angular amplitudes $I_i(q^2)$ are defined in terms of seven transversity amplitudes $A_{\perp,\parallel,0}^{L,R}$ and $A_t$ which are related to the helicity amplitudes by \cite{observables1}
      
      \begin{align}
      A_{\perp,\parallel}&=(H_{+}\mp H_{-})/\sqrt{2};\nonumber \\
      A_0&=H_0 
      \end{align}
      $A_t$ corresponds to the timelike polarization vector of gauge boson which is unphysical and arises because of off-shellness of $K^*$.  
      \begin{equation}
      A_t=\mathcal{M}_{0,t}(B\to K^*\ell^+\ell^-)
      \end{equation}
      There is an additional transversity amplitude, $A_s$ which contributes in NP scenarios where scalar operators are present. 
      Angular coefficients are defined as, 
      \begin{align}
      I_1^s&=\frac{(2+\beta^2_l)}{4}\left[|A_{\perp}^L|^2+|A_{\parallel}^L|^2+(L\rightarrow R)\right]+\frac{4m_l^2}{q^2}Re(A_{\perp}^L A_{\perp}^{R*}+A_{\parallel}^L A_{\parallel}^{R*})\\
      I_1^c&=|A_0^L|^2+|A_0^R|^2+\frac{4 m_l^2}{q^2} \left[|A_t|^2+2 Re(A_0^LA_0^{R*})\right]+\beta_l^2|A_S|^2,\\
      I_2^s&=\frac{\beta_l^2}{4} \left[|A_{\perp}^L|^2+|A_{\parallel}^L|^2+(L\rightarrow R)\right],\\
      I_2^c&= -\beta_l^2\left[|A_0^L|^2+(L\rightarrow R)\right],\\
      I_3&=\frac{1}{2}\beta_l^2 \left[|A_{\perp}^L|^2-|A_{\parallel}^L|^2+(L\rightarrow R)\right],\\
      I_4&=\frac{\beta_l^2}{\sqrt{2}}\left[Re(A_0^LA_{\parallel}^{L*})+(L\rightarrow R)\right],\\
      I_5&=\sqrt{2}\beta_l \left[Re(A_0^LA_{\perp}^{L*})-(L\rightarrow R)-\frac{m_l}{\sqrt{q^2}} Re(A_{\parallel}^L A_S^* + A_{\parallel}^R A_S^*)\right] ,\\
      I_6^s&=2\beta_l\left[Re(A_{\parallel}^LA_{\perp}^{L*})-(L\rightarrow R)\right],\\
      I_6^c&=4\beta_l \frac{m_l}{\sqrt{q^2}}Re\left[A_0^LA_S^*+(L\rightarrow R)\right],\\
      I_7&=\sqrt{2} \beta_l \left[Im(A_0^LA_{\parallel}^*)-(L\rightarrow R)+\frac{m_l}{\sqrt{q^2}}Im(A_{\perp}^LA_S^*+A_{\perp}^RA_S^*)\right],\\
      I_8&=\frac{\beta^2_l}{\sqrt{2}}\left[Im(A_0^LA_{\perp}^{L*})+(L\rightarrow R)\right],\\
      I_9&=\beta_l^2\left[Im(A_{\parallel}^{L*}A_{\perp}^L)+(L\rightarrow R)\right]
      \end{align}
      where the transversity amplitudes are given as, 
      \begin{align}
      A_{\perp}^{L,R}&= \sqrt{2} Nm_B (1-\hat{s})\left[C_9^{eff}\mp C_{10}+2\frac{\hat{m_b}}{\hat{s}}C_7^{eff}\right]\xi_{\perp}(E_{K^*}),\\
      A_{\parallel}^{L,R}& =-\sqrt{2} Nm_B (1-\hat{s})\left[C_9^{eff}\mp C_{10}+2\frac{\hat{m_b}}{\hat{s}}C_7^{eff}\right]\xi_{\perp}(E_{K^*}),\\
      A_0^{L,R}&=-\frac{-Nm_B}{2\hat{m}_{K^*}\sqrt{\hat{s}}} (1-\hat{s})^2\left[C_9^{eff}\mp C_{10}+2\hat{m}_b C_7^{eff}\right]\xi_{\parallel}(E_{K^*}),\\
      A_t& =\frac{Nm_B}{\hat{m}_{K^*}\sqrt{\hat{s}}}(1-\hat{s})^2 C_{10}\xi_{\parallel}(E_{K^*}),\\
      A_S& =\frac{Nm_B^2}{\hat{m}_{K^*}}(1-\hat{s})^2C_S \xi_{\parallel}(E_{K^*})
      \end{align}
      \subsection{Input Parameters}\label{input}
      The Wilson coefficients in Table \ref{inputs} have been taken from \cite{symmetries}.
      \begin{table}[h]
      	\begin{center}
      		\begin{tabular}{|c|c||c|c|} \hline \hline
      			$G_F$ & $1.16 \times 10^{-5} Gev^{-2}$ & $\alpha_{em}$ & 1/137 \\
      			$m_B$ & 5.27 Gev & $m_{K^*}$  &0.896 Gev \\
      			$m_b$ & 4.68 Gev & $m_c$ & 1.4 Gev \\
      			$\lambda$ & 0.22 & A & 0.8 \\
      			$\bar{\rho}$ & 0.19 & $\bar{\eta}$ & 0.36 \\ \hline
      			$C_1$ & -0.257 & $C_2$ & 1.009 \\
      			$C_3$ & -0.005 & $C_4$ & -0.078 \\
      			$C_5$ & 0.0 & $C_6$ & 0.001\\
      			$C_7^{eff}$& -0.304 & $C_8^{eff}$ & -0.1670 \\
      			$C_9$ & 4.211 & $C_{10}$ & -4.103 \\ \hline 
      		\end{tabular}
      	\end{center}
      	\caption{Input parameters}
      	\label{inputs}
      \end{table}
      The form factors have been parameterized in the following way \cite{inputformfactors} :
      \begin{equation}
      F(q^2)=(1-\frac{s}{m_R^2})^{-1}(\alpha_0+\alpha_1(z(q^2)-z(0))+\alpha_2(z(q^2)-z(0))^2)
      \end{equation}
      where $$z(t)= \frac{\sqrt{t_+-t}-\sqrt{t_+-t_0}}{\sqrt{t_+-t}+\sqrt{t_+-t_0}}$$
      where, $t_{\pm}=(m_B\pm m_{K^*})^2$, $t_0=t_+(1-\sqrt{1-t_+/t_-})$, $\alpha_0,\alpha_1,\alpha_2$ are fitting parameters and $m_R$ is the mass of resonance corresponding to quantum number of the form factor (or transition current) for the $b\rightarrow s$ transition. 
      \begin{table}[h]
      	\begin{center}
      		\begin{tabular}{|c|c|c|c|c|}
      			\hline \hline
      			Form Factor & $m_R$ & $\alpha_0$& $\alpha_1$& $\alpha_2$\\ \hline
      			V& 5.415 & $0.34 \pm 0.04$& $-1.05\pm 0.24$ &$ 2.37\pm 1.39$ \\
      			$A_1$& 5.829 &$ 0.27\pm 0.03$& $0.30\pm 0.19$&$ -0.11\pm 0.48$\\ 
      			$A_{12}$& 5.829& $0.26\pm 0.03$&$0.60\pm 0.2$ &$ 0.12\pm 0.84$ \\ \hline
      		\end{tabular}
      	\end{center}
      \end{table}
      $A_2(q^2)$ is defined in terms of other form factors as,
      \begin{equation}
      A_2(q^2)=\frac{\left((m_B+m_{K^*})^2(m_B^2-m_{K^*}^2-q^2)A_1(q^2)-16 m_Bm_{K^*}^2(m_B+m_{K^*}A_{12}(q^2))\right)}{\left((m_B+m_{K^*})^2-q^2\right)\left((m_B-m_{K^*})^2-q^2\right)}
        \end{equation}
    where V, A$_1$ and A$_2$ are three of the seven full form factors \cite{symmetries}. Soft form factors are related to full form factors as, 
 
  \begin{align}
 \xi_{\perp}(q^2)&=\frac{m_B}{m_B+m_{K^*}}V(q^2),\\
 \xi_{\parallel}(q^2)&=\frac{m_B+m_{K^*}}{2 E} A_1(q^2)-\frac{m_B-m_{K^*}}{m_B} A_2(q^2)
 \end{align}
   
      \bibliographystyle{unsrt}
      \bibliography{paper}
\end{document}